\documentclass[aps, prl, twocolumn, letter, superscriptaddress, longbibliography]{revtex4-1}

\newcommand{\mm}{\mathrm}

\usepackage{bm}

\newcommand{\rf}{{\bm f}}
\usepackage{verbatim}

\usepackage[pdftex]{graphicx}

\usepackage{sidecap}
\usepackage[center]{subfigure}

\usepackage{rotating}

\usepackage[usenames]{xcolor}
\usepackage[normalem]{ulem}

\begin{document}

%\title{Near Crystalline Systems Behave Amorphously}
%\title{Gardner Phase in Near Crystalline Systems}
%\title{Gardner Phase in Infinitesimally Frustrated Crystalline Systems}
\title{Glassy, Gardner-like Phenomenology in Minimally Polydisperse Crystalline Systems}
%\title{Glassy Phenomenology in Minimally Polydisperse Crystalline Systems}

\author{Patrick Charbonneau}
\affiliation{Department of Chemistry, Duke University, Durham, North Carolina 27708, USA}
\affiliation{Department of Physics, Duke University, Durham, North Carolina 27708, USA}

\author{Eric I. Corwin}
\affiliation{Department of Physics, University of Oregon, Eugene, Oregon 97403, USA}

\author{Lin Fu}
\affiliation{Department of Chemistry, Duke University, Durham, North Carolina 27708, USA}

\author{Georgios Tsekenis}
\email{geotsek@gmail.com}
\affiliation{Department of Physics, University of Oregon, Eugene, Oregon 97403, USA}

\author{Michael van der Naald}
\affiliation{Department of Physics, University of Oregon, Eugene, Oregon 97403, USA}

\begin{abstract}
We report on a non-equilibrium phase of matter, the minimally disordered crystal phase, which we find exists between the maximally amorphous glasses and the ideal crystal. Even though these near crystals appear highly ordered, they display glassy and jamming features akin to those observed in amorphous solids. Structurally, they exhibit a power-law scaling in their probability distribution of weak forces and small interparticle gaps as well as a flat density of vibrational states.  Dynamically, they display anomalous aging above a characteristic pressure. Quantitatively this disordered crystal phase has much in common with the Gardner-like phase seen in maximally disordered solids. Near crystals should be amenable to experimental realizations in commercially-available particulate systems and are to be indispensable in verifying the theory of amorphous materials. 

%We study the structure and dynamics of crystals of minimally polydisperse hard spheres at high pressures. Structurally, they exhibit a power-law scaling in their probability distribution of weak forces and small interparticle gaps as well as a flat density of vibrational states.  Dynamically, they display anomalous aging beyond a characteristic pressure.  Although essentially crystalline, these solids thus display features reminiscent of the Gardner phase observed in certain amorphous solids. Because preparing these materials is fast and facile, they are ideal for testing a theory of amorphous materials. They are also amenable to experimental realizations in commercially-available particulate systems.

%abstract current word count ???
\end{abstract}

\maketitle

\begin{figure}
\centering
\includegraphics[width=\columnwidth]{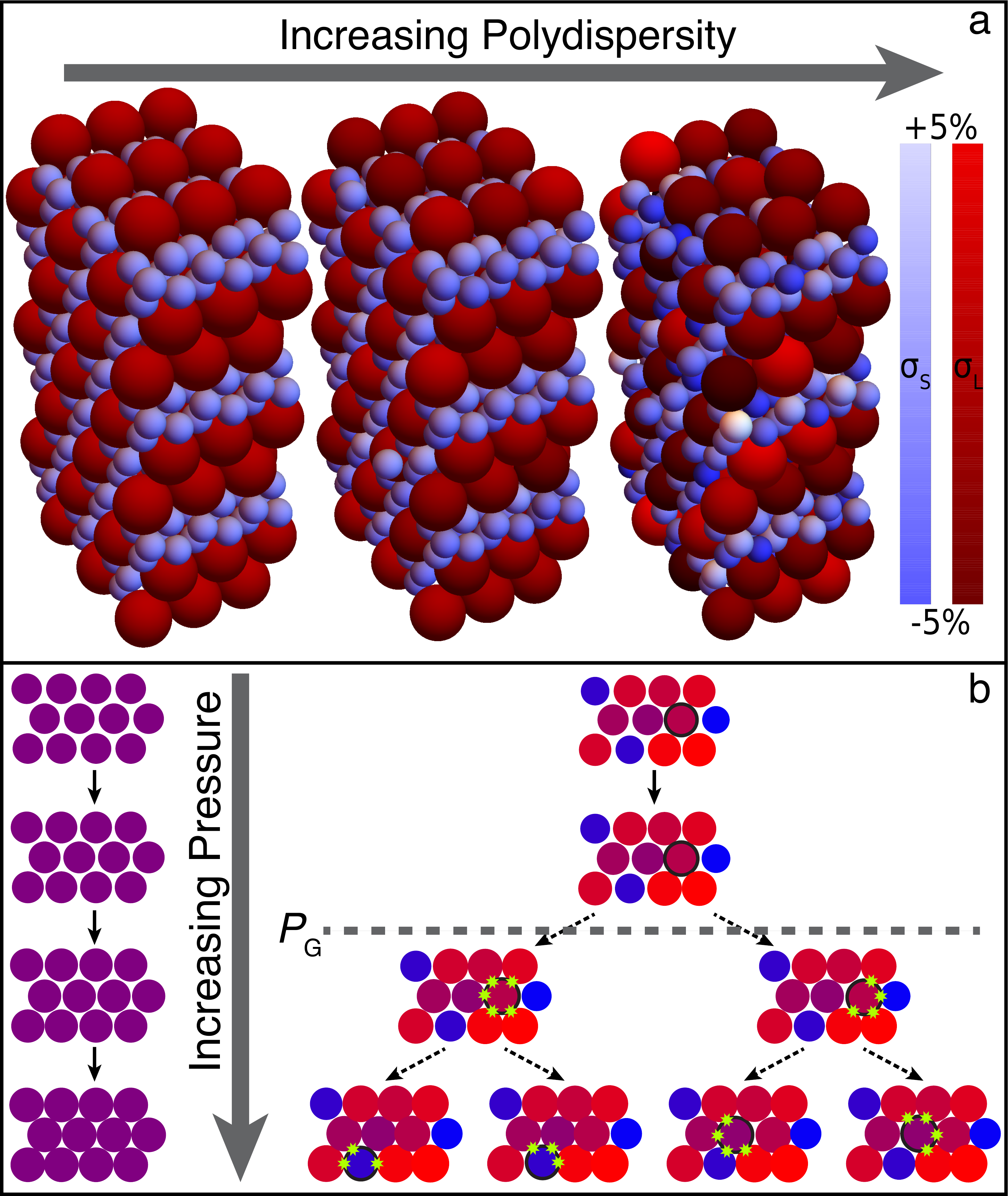}
\caption{(a) Jammed HS1 packings with $s=0.0, 0.01$ and $0.03$, from left to right.  Color encodes the particle diameter, $\sigma_i$. Even the most disordered system appears crystalline.  Note that the unit cell of a perfect HS1 crystal comprises four larger particles and twelve smaller particles, which for a diameter ratio of $1 : 0.5147$ achieves close packing, $\varphi_\mathrm{cp} = 0.7573$. (b) Schematic of a Gardner-like scenario for polydisperse crystals.  While a monodisperse packing has but one well-separated densest packing, the number of nearby optima in a polydisperse system can be large. Beyond a threshold pressure, $P_\mathrm{G}$, constraints on that optimum start to lock in. A particle (outlined in black) is free to collide with all its nearest neighbors at low pressures, but is forced to have one or another set of contacts (green stars) as pressure increases beyond $P_{\mathrm{G}}$.}
\label{fig:visualize}
\end{figure}

\emph{Introduction.--} Supercooling a liquid to form a glass and crunching grains until they jam both lead to solids that are amorphous. Because the two protocols are far out of equilibrium, however, their end products need not have much in common. Twenty years ago, Liu and Nagel nonetheless postulated the existence of a deep connection between them~\cite{liu_nonlinear_1998}, and a formal relationship has recently been uncovered for certain models~\cite{charbonneau_glass_2017}. At the crux of the latter lies the Gardner transition~\cite{gardner_spin_1985, gross_mean-field_1985}, which for a mean-field model of hard spheres is intermediate between glass formation and jamming~\cite{seguin_experimental_2016, berthier_growing_2016, charbonneau_fractal_2014, charbonneau_glass_2017}. At this transition, the phase space of a mechanically stable glass basin splits into an intricate and hierarchical arrangement of marginally stable sub-basins; jamming occurs deep within this marginal phase. %experiments to cite \cite{gebbia_glassy_2017}
Remarkably, mean-field theory (MFT) further predicts materials features that are robustly universal down to dimension $d=2$~\cite{charbonneau_glass_2017}. For instance, amorphous packings of hard spheres exhibit distinctive power-law distributed small interparticle gaps and weak contact forces with exponents that are numerically consistent with MFT~\cite{skoge_packing_2006, wyart_marginal_2012, lerner_low-energy_2013, charbonneau_universal_2012, charbonneau_jamming_2015, charbonneau_glass_2017}. 
A similarly stunning agreement is observed for the distribution of vibrational excitations at and around these jammed configurations~\cite{ohern_random_2002, ohern_jamming_2003, wyart_effects_2005, charbonneau_universal_2016}. %MFT, however, only describes solids that are fully amorphous.

While the description of crystalline solids has long been well established and that of amorphous solids is under increasingly strong theoretical control, a large conceptual gap persists in between these two materials poles. Various proposals to reconcile them have recently emerged. Goodrich et al.~found that athermal crystals with discrete disorder, such as vacancies and interstitials, display structural and rheological properties similar to those of amorphous solids~\cite{goodrich_solids_2014}.  Such crystals also undergo a relatively sharp amorphization transition as the particle size dispersity (polydispersity) increases~\cite{mizuno_elastic_2013, tong_crystals_2015}. For jammed packings specifically, Tong et al.~proposed that a disordered crystal phase underlies distinct scaling exponents for certain rheological quantities, such as the ratio of the shear to bulk modulus~\cite{tong_crystals_2015}. The microscopic origin of these anomalies in slightly disordered crystals, however, remains far from understood. 

In this Letter, we investigate the out-of-equilibrium physics of crystals of weakly polydisperse particles. Disorder is introduced continuously in otherwise perfect crystals of hard spheres by scaling particle radii by a factor drawn from a log-normal distribution of unit mean and standard deviation $s$~\cite{foot:frac}. The chosen crystal symmetry, HS1~\cite{otoole_new_2011} (\cite[Sect.~II]{SI}), contains no particle with coplanar neighbors -- unlike face-centered cubic (FCC) and many other crystal symmetries -- hence the role of low-energy buckling excitations is minimal~\cite{charbonneau_jamming_2015}. We study both the relaxation dynamics of finite-pressure crystals and the structure of infinite-pressure jammed packings. 
Remarkably, even though these solids appear crystalline (Fig.~\ref{fig:visualize}), we find that their structure and dynamics exhibit most of the glassy properties of amorphous solids, in line with the MFT predictions for high-density amorphous solids. %of the Gardner phase.

\begin{figure*}[ht]
\centering
\includegraphics[width=\textwidth]{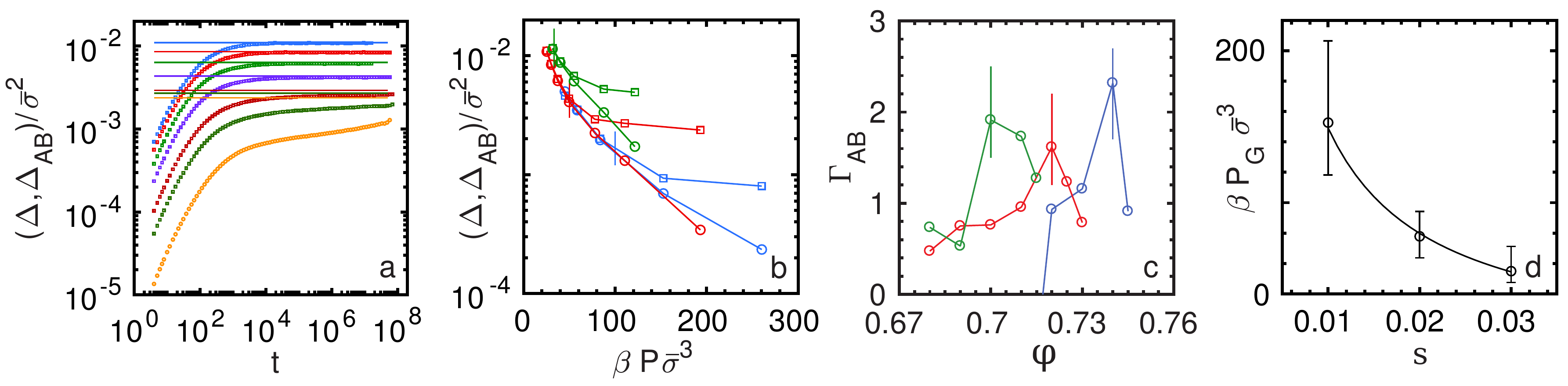}
\caption{(a) $\Delta(t, 0)$ (squares) and $\Delta_{AB}(t) \equiv \Delta(\infty, 0)$ (lines) for HS1 crystals with $s=0.02$. As $\varphi$ increases (from top to bottom), $\Delta(t,0)$ crosses over from having a well-defined long-time plateau to displaying logarithmic aging at $\varphi_{\mm G} \approx 0.72$.
(b) Evolution of $\Delta_{AB}$ and (early) plateau height of $\Delta(t, t_{\mm w})$ with pressure for $s=0.01$ (blue), $0.02$ (red) and $0.03$ (green).
(c) The skewness, $\Gamma_{\mm AB}$, of the distributions of $\Delta_{\mm AB}$ for each polydispersity peaks at $\varphi_{\mm G}$ denoted with vertical lines in (c) which in turn defines $P_{\mm G}$ denoted with vertical lines in (b).
(d) The pressure, $P_\mathrm{G}$, corresponding to $\varphi_{\mm G}$ increases with decreasing polydispersity.  The solid line is a fit to an inverse relationship, which suggests that the anomalous regime only vanishes for $s\rightarrow0$, where $P_\mathrm{G}\rightarrow\infty$. For the sake of comparison, in previous works (with uniformly distributed polydispersity), equilibrium polydisperse FCC crystals become unstable to fractionation around $s\sim0.08$~\cite{sollich_crystalline_2010, sollich_polydispersity_2011}, and the athermal amorphization transition occurs around $s\sim0.11$ \cite{tong_crystals_2015}.
}
\label{fig:Deltas}
\end{figure*}

\emph{Glassy Dynamics.--} We probe the dynamics of $300-400$ copies of systems with $N=2000$ particles initialized near the melting density of the HS1 lattice and annealed following a standard protocol~\cite{berthier_growing_2016}. First, we run isothermal-isobaric, constant $NPT$, Monte Carlo (MC) simulations using a relatively high pressure quench, until a target packing fraction, $\varphi$, is reached. Isothermal-isochoric, constant $NVT$, Monte Carlo simulations are then run using only local particle displacements (\cite[Sect.~IIIA]{SI}).
The roughness of the caging landscape is ascertained by the long-time behavior of the mean-squared displacement of the particle positions, $\vec{r}$,
\begin{equation}
\Delta(t,t_{\mm w}) = \frac{1}{N} \sum_{i=1}^{N}\left<| \vec{r}_{i}(t+t_{\mm w})-\vec{r}_{i}(t_{\mm w})|^2\right>,
\end{equation}
where $t_{\mm w}$ is the time (measured in sweeps of $N$ MC steps) after reaching a target $\varphi$.  For a simple, mechanically stable thermal solid, $\Delta(t, t_{\mm w})$ is expected to plateau quickly because all particles can efficiently sample their local cage. For a marginally stable solid, by contrast, $\Delta(t, t_{\mm w})$ is expected to exhibit significant aging, a reflection of the difficulty of sampling the complex caging landscape associated with this regime~\cite{berthier_growing_2016}. In the latter case, the long-time limit of $\Delta(t, t_{\mm w})$ is computationally out of reach, even for the relatively small systems studied here. We thus also compute the distance between two system copies, $A$ and $B$,
\begin{equation}
\Delta_{AB} = \frac{1}{N} \sum_{i=1}^{N}\left<|\vec{r}_{i}^{A}(t)-\vec{r}_{i}^{B}(t)|^2\right>, \forall t
\end{equation}
with the same $\varphi$ and quenched disorder, but evolved from different stochastic trajectories, such that $\Delta_{AB}=\Delta(t\rightarrow\infty, t_{\mm w})$. %Note that the two crystal centers of mass are kept aligned in order to eliminate spurious contributions due to drift.

Figure~\ref{fig:Deltas}a shows that aging, which is undetectable at low pressures, first appears and then becomes increasingly notable as pressure increases. The early plateau of $\Delta(t,t_{\mm{w}})$ correspondingly splits from $\Delta_{AB}$ (Fig.~\ref{fig:Deltas}b~\cite[Sect.~IIIA]{SI}). As in Ref.~\onlinecite{berthier_growing_2016}, the skewness, $\Gamma_{AB}$, of the distribution of $\Delta_{AB}$ for different initial configurations also peaks in that regime, which provides a clear definition of $\varphi_\mathrm{G}$ (Fig.~\ref{fig:Deltas}c). Both effects are akin to the anomalous phenomenology observed in glassy hard spheres at high pressure~\cite{berthier_growing_2016}.  Remarkably, as $s$ decreases, the onset of aging and $\varphi_\mathrm{G}$, are both pushed to increasingly larger pressures (Fig.~\ref{fig:Deltas}d), while the equation of state is barely affected (\cite[Sect.~IIIA]{SI}).  Microscopically, the Gardner-like regime appears when the typical interparticle spacing, which scales as $1/P$, becomes comparable to the polydispersity, i.e., $P_\mathrm{G} \sim 1/s$ (Fig.~\ref{fig:Deltas}d).  The anomalous regime thus only disappears for a perfect crystal, i.e., for $s\rightarrow 0$. This effect is reminiscent of the Gardner regime of amorphous hard spheres, which also steadily shrinks as the ideal glass limit is approached~\cite{charbonneau_glass_2017}. Although computer simulations, as considered here, do not cover the thermodynamic limit to determine whether a true phase transition takes place, our observations are thus consistent with the Gardner-like regime observed in numerical studies of hard-sphere glasses~\cite{berthier_growing_2016}. 

\emph{Isostatic Mechanical Equilibrium.--} Having established that polydisperse hard sphere crystals display anomalous features at high but still finite pressure, we compare their micro-structures at infinite pressure (jamming) with those of amorphous jammed configurations. Jammed packings of $N=432$ polydisperse soft spheres in HS1 symmetry are obtained by minimizing the energy of $466-736$ realizations for each $s$ studied~\cite{charbonneau_universal_2012,charbonneau_numerical_2015} (\cite[Sect.~IIIB]{SI}). (For $s\lesssim0.01$, the unambiguous detection of small forces and gaps near the numerical accuracy of the simulation is prohibitively cumbersome.) The final configurations therefore coincide with the inherent structures of the polydisperse hard sphere crystals. Just like amorphous jammed packings, these near-crystalline configurations contain but a small fraction of rattling particles and are otherwise perfectly isostatic. The interparticle forces, $f$, can thus be determined directly from the contact vectors~\cite{charbonneau_jamming_2015}.

\begin{figure*}[ht]
\centering
\includegraphics[width=\textwidth]{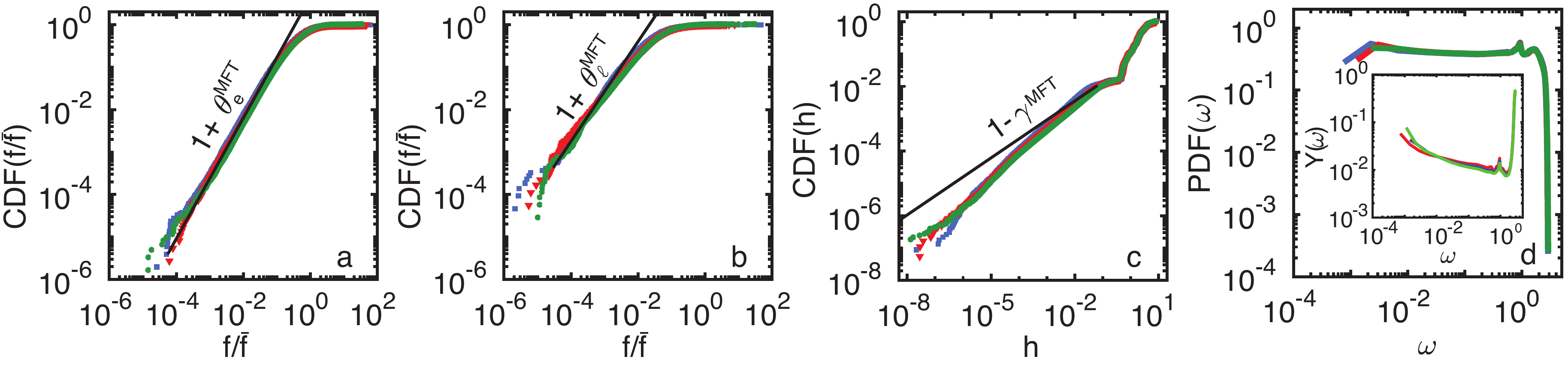}
\caption{Cumulative distribution function, $\mm{CDF}(x)=\int_0^x{\mm{PDF}(x')dx'}$, of contact forces between pairs of particles associated with (a) extended floppy modes and (b) localized floppy modes for  $s=0.01$ (blue squares), $0.02$ (red triangles) and $0.03$ (green circles). (c) CDF for small interparticle gaps in the same systems. MFT predictions for the power-law exponents, $1+\theta_{\mm e}^{\mathrm{MFT}} = 1.42311$, $1+\theta_{\ell}^{\mathrm{MFT}} = 1.17462$ and $1-\gamma^{\mathrm{MFT}} = 1-0.41269$, are given as black solid lines. While close agreement in observed in (a) and (b), a significant discrepancy is seen in (c). (d) The probability distribution of the frequency of harmonic vibrations has a spectrum identical to that of a disordered jammed packing for all polydispersities, while a standard Debye scaling would have $\sim \omega^{d-1}$. (inset) Evolution of the average IPR with frequency.  Low-frequency modes tend to be quasi-localized, as are those of fully amorphous solids. By contrast, at high frequency both the spectra and the IPR display crystal peaks.}
\label{fig:cdfForcesGaps}
\label{fig:cdfDOSnIPR}
\end{figure*}

Like their amorphous counterparts, our packings have power-law distributed small forces with different scaling exponent for contacts that give rise to localized excitations when opened and those associated with extended excitations \cite{charbonneau_jamming_2015, wyart_marginal_2012, lerner_low-energy_2013}~\cite[Sect.~IIIB]{SI}, i.e.,
\begin{eqnarray} 
\label{eqn:forces}
\mm{PDF}_{\mathrm{e}}(f) \sim f^{\theta_{\mathrm{e}}} \;\;\;\mathrm{and}\;\;\;
\mm{PDF}_{\ell}(f) \sim f^{\theta_{\ell}} ,
\end{eqnarray}
respectively. Figures \ref{fig:cdfForcesGaps}a and \ref{fig:cdfForcesGaps}b reveal that the force scaling exponents are in good agreement with the MFT predictions, $\theta_{\mm e}^{\mathrm{MFT}} = 0.42311$ and $\theta_{\ell}^{\mathrm{MFT}} = 0.17462$. The distribution of interparticle gaps, $h=\frac{r_{ij}}{(\sigma_{i}+\sigma_j)/2}-1$, which is complementary to that of the forces~\cite{wyart_marginal_2012, lerner_low-energy_2013, charbonneau_universal_2012}, also displays a power-law tail
\begin{eqnarray} 
\label{eqn:gap}
\mm{PDF}_h(h) \sim h^{-\gamma},
\end{eqnarray}
(Fig.~\ref{fig:cdfForcesGaps}c). The observed exponent, however, is visibly smaller than the MFT prediction, $\gamma^{\mathrm{MFT}} = 0.41269$, for all $s$ considered. For the range of very small polydispersities considered we nonetheless clearly observe that near-crystals have a complex particle microstructure concordant with that of amorphous solids. 

The theory of marginally stable packings provides inequalities for these exponents~\cite{wyart_marginal_2012, lerner_low-energy_2013,muller_marginal_2015}, $\gamma \geq 1/(2+\theta_{\mm e})$ and $\gamma \geq (1-\theta_{\ell})/2$, which were found to be saturated in amorphous solids~\cite{charbonneau_jamming_2015,charbonneau_glass_2017}.  Here, because the force scaling exponents are consistent with the MFT predictions while $\gamma$ is markedly smaller, both inequalities are violated. Even though the treatment in Refs.~\cite{wyart_marginal_2012, lerner_low-energy_2013,muller_marginal_2015} is seemingly independent of the degree of disorder, it implicitly assumes that the marginal solids have no structural correlations. While this may be a reasonably valid assumption for amorphous packings, it is clearly not the case here. How to include such correlations in the theory of marginality and what precise values should the critical exponents take in that context, however, remain open problems. %Another possible resolution of this conundrum is that the gap distribution might inadvertently blend two (or more) distinct microscopic processes and thus an effective exponent could hide the proper $\gamma$. Were that to be the case, orders of magnitude increase in numerical accuracy might be needed to resolve the proper small-gap behavior. From a theoretical standpoint, however, the origin of such putative microscopic distinction remains elusive. 

\emph{Harmonic excitations.--}
As a further test of the similarity between polydisperse crystals and amorphous solids, we consider the low-energy excitations around the jammed minima~\cite{charbonneau_universal_2016, ohern_jamming_2003}. The eigenvalues $\lambda_k$ and eigenvectors $\{\vec{u}_{i}\}_k$ of the Hessian computed from the contact vectors provide the harmonic frequencies, $\omega_k=\sqrt{\lambda_k}$, and normal modes, respectively. As in amorphous solids, we find the spectra of vibrational states to be flat at low frequencies (Fig.~\ref{fig:cdfDOSnIPR}d), and the spatial extent of the normal modes to be nontrivial (Fig.~\ref{fig:cdfDOSnIPR}d, inset). The eigenmodes, $\{\vec{u}_i(\omega_k)\}_k$, at a given $\omega_k$ indeed have an inverse participation ratio (IPR)
\begin{eqnarray} 
\label{ipr}
Y(\omega) = \frac{\sum_i^N | \vec{u}_i(\omega)|^4}{[\sum_i^N | \vec{u}_i(\omega)|^2]^2},
\end{eqnarray}
consistent with them being mostly delocalized at intermediate frequencies with some degree of quasi-localization at low frequencies~\cite{widmer-cooper_irreversible_2008, xu_anharmonic_2010, mizuno_elastic_2013, manning_vibrational_2011} (\cite[Sect.~IIIB]{SI}).
Remarkably, the high-frequency localized peaks of the crystal structure are also preserved. Because a similar normal mode distribution was observed in slightly disordered FCC packings~\cite{mari_jamming_2009}, the density of vibrational states is likely universal in marginally stable packings.

\emph{Conclusion.--}
Our work evinces that minuscule amounts of disorder are sufficient to blend the physics of crystals with that of amorphous solids. Perfect crystalline ground states are therefore a singular limit. %We find that such crystals enter into a state of Gardner phenomenology above a pressure $P_\mathrm{G} \sim 1/s$.  Thus, for any finite polydispersity, no matter how small, there exists a pressure above which the Gardner phenomenology will obtain.  
Because relating microscopic features with macroscopic rheology is still unsolved, it is unclear whether our findings relate with those of the universality class proposed in Ref.~\onlinecite{tong_crystals_2015}, but this hypothesis deserves further consideration. The specific exponent values and their violation of the stability bounds for marginal solids observed in these systems should also motivate additional study. %It remains an open question if novel marginal stability inequalities exist or if there are more microscopic processes at play leading to additional more complicated inequalities. % point for which any amount of polydispersity will cause the system to eventually move into the universality class of amorphous packings.

%\ericemph{I suggest that we remove the following two sentences.  I don't think that the Tong work is entirely relevant and the Goodrich work is a characteristically different thing.  If we are to include it, I'd like to say something along the lines of Goodrich's work interpolates from the crystal to the amorphous by creating local defects and so it's not surprising that a system composed of large crystallites behaving like a crystal.}
%\st{This thermal finding of ours stands in corresponding agreement with work in athermal jammed disordered crystals. Together, our result and the result of~\cite{tong_crystals_2015}, are in contrast with the results found with discrete disorder which appears to assess the crossover from crystalline to amorphous rigidity at a finite amount of disorder~\cite{goodrich_solids_2014}. }

%the 	 packings exhibit force and gap distributions with power law tails at small values.  The exponents associated with force scaling are in good correspondence with the MFT predictions while the exponent for the gap distribution is significantly smaller than the MFT prediction.  

The many structural and dynamical similitudes between crystals of polydisperse spheres and amorphous solids suggest that the former could be used to better understand the latter. The simplicity and stability of polydisperse crystals make them ideal for exploring the MFT Gardner transition scenario. Resolving whether a thermodynamic transition exists in finite-dimension~\cite{CharbonneauYaida2017,hicks:2017,Wang:2017,lubchenko_aging_2017,Angelini:2017} and for what interaction types~\cite{scalliet_absence_2017}, in particular, are of acute interest. In practice, commercially manufactured colloids and ball bearings have nominal polydispersities on the order of or larger than that studied here. Such easily accessible experimental systems could thus also be investigated %for the physics they are described by, amorphous versus crystalline, corroborate  our results and 
to expand our understanding of rigidity in the entire spectrum from perfect order to maximal disorder~\cite{gebbia_glassy_2017}. % (for a recent experimental work in minimally disordered crystals see).

%(which has been harder to detect in structural glass formers with soft-sphere interactions)

\emph{Acknowledgements:}
We thank L.~Berthier, E.~Lerner, P.~Urbani, M.~Wyart and F.~Zamponi for useful discussions.  
We also thank G.~Parisi for providing the original impetus to explore the behavior of Gardner phenomena in a perturbed lattice.
This work benefited from access to the University of Oregon high performance computer, Talapas.
Simulations were also run on the Duke (University) Computer Cluster.
We gratefully acknowledge the support of NVIDIA Corporation with the donation of a Titan X Pascal GPU used in part for this research.
This work was supported by Grants from the Simons Foundation (No. 454937, P. C.; No. 454939, E. I. C, G. T., M.v.d.N.). 
E.I.C. was supported by the National Science Foundation under Career Grant No. DMR-1255370.
L. F. was supported by the National Science Foundation’s (NSF) Research Triangle Materials Research Science and Engineering Center (MRSEC) under Grant No. (DMR-1121107) and by NSF’s grant from the Nanomanufacturing Program (CMMI-1363483).

%\bibliographystyle{apsrev4-1}

%\bibliography{four}

%\end{document}

\setcounter{figure}{0}

\section{Supplementary Material for: Glassy, Gardner-like Phenomenology in Minimally Polydisperse Crystalline Systems}

This Supplementary Material details the crystal structure and the choice of polydispersity as well as the thermal hard sphere simulation and athermal soft spheres energy minimization schemes.

\section{Polydisperse Crystal}

The binary crystal studied in this work is based on the Hudson Structure One (HS1)~\cite{otoole_new_2011}, whose unit cell contains four larger particles and 12 smaller particles. It has orthorhombic periodicity with dimensions $a:b:c=1:1.4980:2.6014$ and for a ratio of smaller to larger particle diameter $\sigma_\mathrm{S}/\sigma_\mathrm{L}\doteq0.5147$ the HS1 crystal attains close packing with $\varphi_\mathrm{cp} \doteq 0.7573$.

In order to introduce polydispersity in this crystal, the particle diameter ($\sigma_\mathrm{L}$ or $\sigma_{\mathrm{S}}$) of each particle is rescaled
\begin{eqnarray*}
\sigma_i = \sigma_\mathrm{L/S} \times R,
\end{eqnarray*}
where $R$ is a log-normal distributed random variable with unit mean and standard deviation $s$. This choice of distribution is fairly generic and avoids the generation of negative diameters. 

\section{Simulation Methods}

\subsection{Gardner Phenomenology}

Simulations are initialized from a perfectly ordered HS1 binary crystal with a lattice spacing just large enough for the overlaps resulting from the instance of polydispersity to be eliminated. Isothermal-isobaric, constant $NPT$, Monte Carlo (MC) simulations are then run to reach a target $\varphi$. Pressure $P$ is kept constant by standard logarithmically-sampled volume moves. Because the initial configurations are well-ordered, conventional MC moves with a ratio between particle moves and volume moves being $N:10$ ($N$ being the number of particles) suffice to efficiently compress the system. Once the target density is reached, constant $NVT$ simulations are performed using a local Metropolis dynamics. Step sizes of the different MC moves are tuned to ensure that the acceptance ratio stays between 40\% and 50\%. 

\subsubsection{Mean-Squared Displacement}
A standard order parameter for glassiness is the plateau height of the mean-squared displacement of particles,
\begin{eqnarray*}
\Delta(t,t_{\mm w}) = \frac{1}{N} \sum_{i=1}^{N}\left<| \vec{r}_{i}(t+t_{\mm w})-\vec{r}_{i}(t_{\mm w})|^2\right>,
\end{eqnarray*}
where $t_{\mm w}$ is the waiting time after the target pressure or density is reached and $\vec{r}_i$ is the position of particle $i$. We measure the early plateau height of $\Delta(t, t_{\mm w}=0)$ for different $\varphi$. For $\varphi\lesssim\varphi_\mathrm{G}$, the early plateau height can be easily estimated because the $\Delta(t, 0)$ quickly reaches a well-defined constant. For $\varphi\lesssim\varphi_\mathrm{G}$, however, a logarithmic aging $\Delta(t, 0) \sim \ln(t)$ quickly develops. In order to estimate the early plateau height, we fit the MSD beyond the transient with $\ln[\Delta(t, 0)] = Q_{1} \ln(t) + Q_{2}$. The early $\Delta(t,0)$ is taken to be the intercept of this fit at $t=1$, i.e., $\Delta(t=1,0)=Q_{2}$. Note that this procedure generalizes naturally to systems with a well-defined plateau.

The long-time limit of $\Delta(t, t_{\mm w})$ quickly becomes computationally unattainable once $\varphi\gtrsim\varphi_\mathrm{G}$. To more clearly reveal the effect of aging, we obtain the equilibrium $\Delta(t\rightarrow\infty, t_{\mm w}\rightarrow\infty)$ from the distance $\Delta_{AB}$ between two different copies, $A$ and $B$, with the same $\varphi$ and particle polydispersity, compressed from the same initial configuration, but using a different stochastic trajectory
\begin{eqnarray*}
\Delta_{AB}(t) = \frac{1}{N} \sum_{i=1}^{N}\left<|\vec{r}_{i}^{A}(t)-\vec{r}_{i}^{B}(t)|^2\right>.
\end{eqnarray*}
Note that $\Delta_{AB}(t)$ is calculated after aligning the centers of mass of the two copies. A few hundred realizations of disorder are used in the averaging for both $\Delta(t, t_{\mm w})$ and $\Delta_{AB}$. 

Introducing polydispersity changes particle sizes non-uniformly, and thus a finite-size system cannot reach its densest packing while maintaining the original aspect ratio of the simulation box. Even though the $s$ considered here and the resulting cell anisotropy are very small, we employ anisotropic volume moves to compute $\Delta(t,t_\mathrm{w})$. For $\Delta_{AB}(t)$, however, only isotropic volume moves are used to ensure that independent system copies have the same dimensions.

\subsubsection{Equation of State}

The system pressure is calculated from the virial equation of state (EoS). In general, for a polydisperse system, this would require calculating of $N(N-1)/2$ distinct pair distribution functions. For hard interactions, however, a rescaling reduces the relationship to a single distribution function. Defining the rescaled quantities:
\begin{eqnarray*}
	\bar{r}_{ij} &=& \frac{r_{ij}}{\sigma_{ij}}\\
   	\bar{u}_{ij} & = & \left \{
	\begin{array}{lr}
		\infty, &  \bar{r}_{ij} < 1 \\ 
       		0,  &   \bar{r}_{ij} \geq 1
	\end{array}
	\right.\\
	\bar{\rf}_{ij} &=& -\nabla \bar{u}_{ij}
\end{eqnarray*}
where $r_{ij}=|\vec{r}_i - \vec{r}_j|$ is the distance between particles $i$ and $j$, $\sigma_{ij}=(\sigma_i+\sigma_j)/2$, $\sigma_i$ is the diameter of particle $i$, we can indeed rewrite the virial as 
\begin{eqnarray*}
\beta P 
&=& \rho  + \frac{\beta}{3V}\left<\sum_{i<j}\rf_{ij}\cdot\vec{r}_{ij}\right>\\
&=& \rho + \frac{\beta}{3V}\sum_{i<j}\left<\rf_{ij}\cdot\vec{r}_{ij}\right>\\
&=& \rho + \frac{\beta\rho^2}{3}\frac{1}{N(N - 1)}\sum_{i<j}\int \rf_{ij}\cdot\vec{r}_{ij}g(\vec{r}_{ij})d\vec{r}_{ij}\\
&=& \rho + \frac{4\beta\pi\rho^2}{3}\frac{1}{N(N - 1)} \sum_{i<j}\sigma_{ij}^3\int \bar{f}_{ij}\bar{r}_{ij}^3\bar{g}(\bar{r}_{ij})d\bar{r}_{ij}\\
&=& \rho + \frac{4\pi\rho^2}{3}\bar{g}(1^+)\sum_{i<j}\frac{\sigma_{ij}^3}{N(N - 1)},
\label{eqn:virial}
\end{eqnarray*}
where $\bar{g}(\bar{r})$ is a uniform rescaled pair distribution function~\cite{santos_contact_2002} and the contact value $\bar{g}(1^+)$ is extrapolated from the first few non-zero values of $\bar{g}(\bar{r})$ using a quadratic fit. Note that the above expression reduces to the monodisperse case when $\sigma_i=\sigma_j=\sigma,\ \ \forall i,j$.

Once a target $\varphi$ is reached, $2\times10^5$ MC cycles are first run to equilibrate the system. The distribution function $\bar{g}(\bar{r})$ is then sampled every 100 MC cycles, and $\bar{g}(1+)$ is updated after each sampling. Except for the few largest $\varphi$, $P$ quickly converges to its equilibrium value. For the few largest $\varphi$, the last recorded (out-of-equilibrium) value of $P$ is reported.

\begin{figure}[th]
\renewcommand{\thefigure}{S\arabic{figure}}
\centering
\includegraphics[scale=.45]{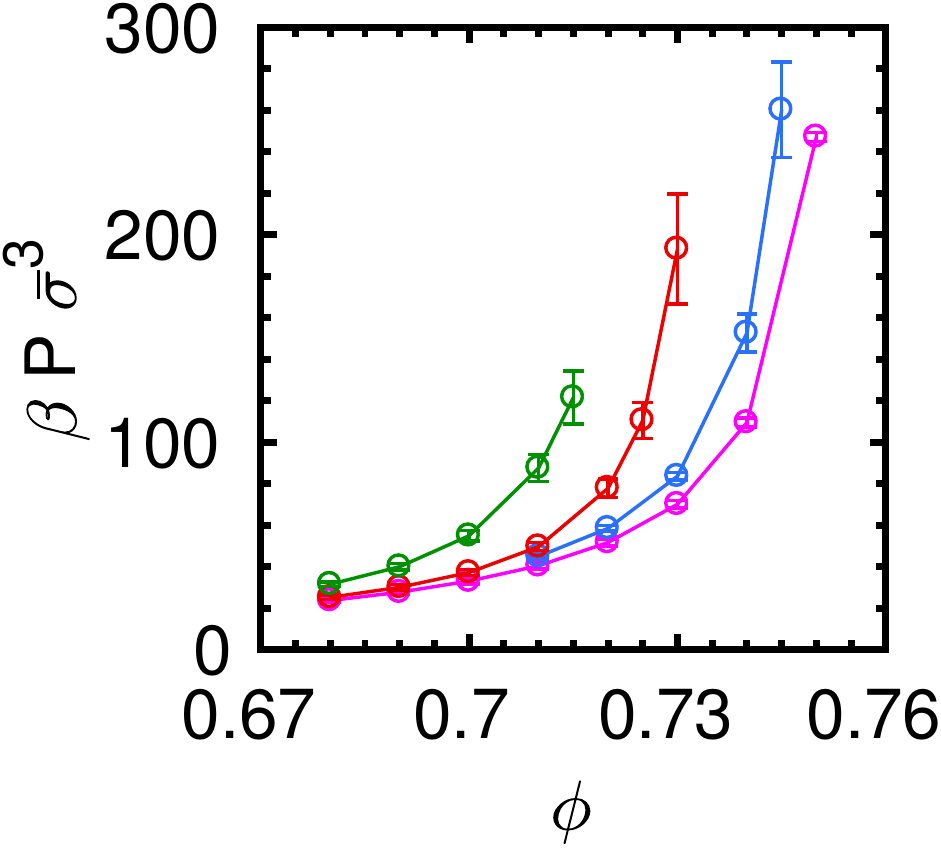}
\includegraphics[scale=.45]{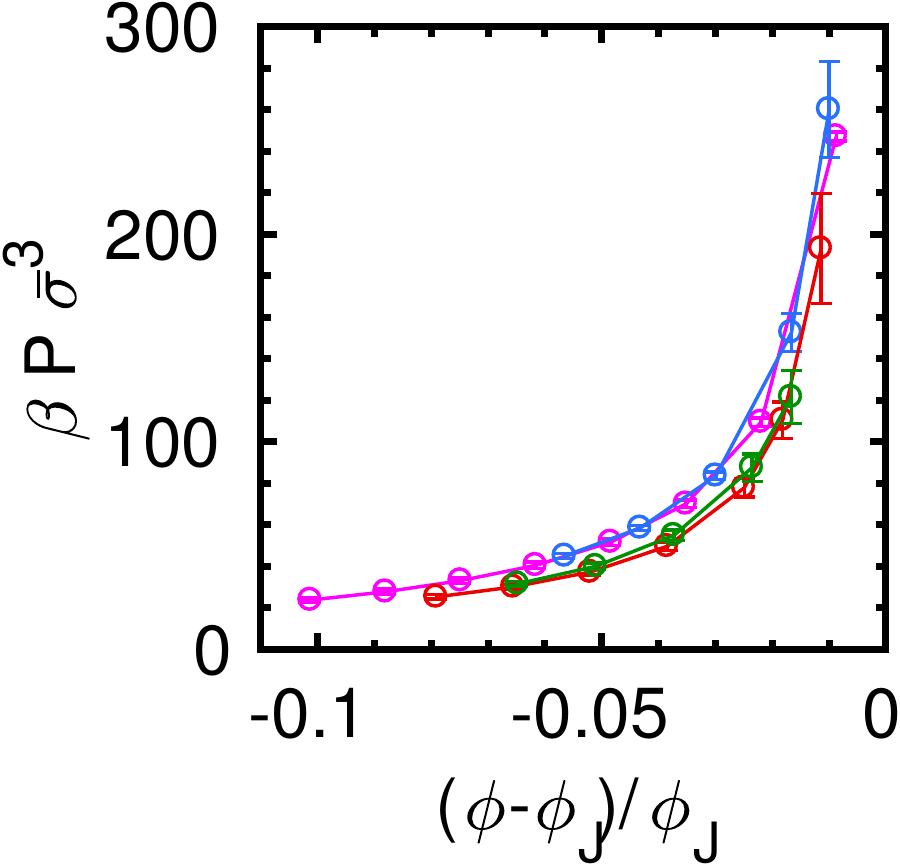}
\caption{
The equation of state (left panel) for different polydispersities (from left to right), $s=0.03$ (green), $s=0.02$ (red), $s=0.01$ (blue), $s=0$ (purple). (right panel) Collapse of the EoS after rescaling with the jamming packing fraction $\varphi_{\mathrm{J}}$.
}
\label{fig:betaPvsPhi}
\end{figure}

Pressure diverges at a finite packing fraction as can be seen in Figure~\ref{fig:betaPvsPhi}.  Towards infinite pressure the system indeed appears to be asymptoting to its jamming behavior at $\varphi_{J}$. A collapse is thus obtained after rescaling the packing fraction with the distance to jamming (Figure~\ref{fig:betaPvsPhi}, right panel).

\subsection{Inherent Structures}

\subsubsection{Athermal Energy Minimization Method}
We prepare HS1 packings of harmonic soft spheres at jamming, starting from a packing fraction well above the crystal density, and then successively minimize the energy and shrink the particles until overlaps between spheres have all vanished. The preparation scheme follows that of Refs.~\onlinecite{charbonneau_jamming_2015,charbonneau_universal_2016} with initial packing fraction $\varphi_\mathrm{i} = 0.8$ and convergence criterion that the packing (with rattlers removed) reaches isostatic equilibrium with $N_{\mathrm{c}}=(N-1)d+1$ contacts, as is expected for a system under periodic boundary conditions~\cite{charbonneau_jamming_2015}. 

\subsubsection{Force Network Calculation}
Interparticle forces in isostatic packings can be uniquely determined from the contact vectors  (see e.g., Ref.~\onlinecite{charbonneau_jamming_2015}).  Before calculating the forces, we remove all rattlers, which are particles that with fewer than $d+1$ contacts or with contacts that are co-hemispheric. 

\subsubsection{Extended and Localized Floppy Modes}
In order to determine whether a contact is associated with an extended or a localized floppy mode, we follow the scheme described in Ref.~\onlinecite{charbonneau_jamming_2015} to extract the particle displacements in response to opening a contact.

More specifically, we solve for
\begin{eqnarray*} 
	H \delta \vec{r}^{(\tau)} = S^T \vec{\tau}
\end{eqnarray*}
where $S$ is the contact matrix, $H=S^T S$ is the Hessian of the packing and $\vec{\tau}=\delta_{\tau,<kl>}$ is a vector containing a unit entry at contact $\tau$ with all other contacts $<kl>$ zeroed. The solution of this equation gives the particle displacements, $\delta\vec{r}^{(\tau)}$, associated with opening contact $\tau$. 

A singular value decomposition of a non-square matrix can generally be expressed as $S=U\Sigma V^T$, where $\Sigma$ is the rectangular diagonal matrix with the singular values (non-negative real numbers) in its diagonal, and $U$ and $V$ are the square matrices of the left-singular and right-singular eigenvectors. We can invert $H$ by using only the non-zero singular values and the corresponding left and right eigenvectors of the contact matrix $S^T$. We then obtain
\begin{eqnarray*} 
	\delta \vec{r}^{(\tau)} = V \Sigma^{-1} U^T \vec{\tau},
\label{eq666}
\end{eqnarray*}
which can be solved iteratively for each $\tau$.  The floppy modes fall naturally into two categories according to the relative value of their median, $V_{\mathrm{median}}=\mathrm{median}\{\delta r_i \}$, to their mean, $V_{\mathrm{mean}}=\mathrm{mean}\{\delta r_i \}$, displacements. Extended and localized modes are characterized by high and low ratios of $V_{\mathrm{median}}/V_{\mathrm{mean}}$ respectively with a split naturally occurring between them at $V_{\mathrm{median}}/V_{\mathrm{mean}}=0.1$. 
%\pc{it would be good to think of depositing these codes along with the data for this manuscript.}

\subsubsection{Gap Distribution}
The gap between pairs of particles, $h = r_{ij}/\sigma_{ij} - 1$, is computed neglecting rattlers and pairs of particles already in contact. 

\subsubsection{Vibrational States}
The vibrational states of the packing are obtained from the Hessian, which is computed as in Ref.~\cite{charbonneau_universal_2016}. We diagonalize $H$ to compute its eigenvalues $\lambda_k$, and thus the vibrational frequencies, $\omega_k=\sqrt{\lambda_k}$, of the normal modes of the configuration. The inverse participation ratio (IPR) of the associated eigenvectors, $\{\vec{u}_i(\omega_k)\}_k$, provides a measure of the spatial extent of the normal modes. Note that because we are only interested in the spectrum of the mechanically rigid portion of the packing modes associated with rattlers are removed from the analysis. 

\bibliography{four}

\end{document}